\documentclass[aps,prl,twocolumn,showpacs]{revtex4}
\usepackage{amsfonts}
\usepackage{amsmath}
\usepackage{amssymb}
\usepackage{graphicx}
\usepackage{wasysym}

\def\kbar{\protect\@kbar}
\def\@kbar{\relax \bgroup
\def\@tempa{\hbox{\raise.73\ht0
\hbox to0pt{\kern.25\wd0\vrule width.5\wd0 height.1pt
depth.1pt\hss}\box0}}\mathchoice{\setbox0\hbox{$\displaystyle
k$}\@tempa}{\setbox0\hbox{$\textstyle
k$}\@tempa}{\setbox0\hbox{$\scriptstyle
k$}\@tempa}{\setbox0\hbox{$\scriptscriptstyle k$}\@tempa}\egroup}

\begin{document}

\title{\textbf{Topological Properties of Adiabatically Varied Floquet Systems}}
\author{Itzhack Dana}
\affiliation{Minerva Center and Department of Physics, Bar-Ilan University, Ramat-Gan
52900, Israel}

\begin{abstract}
Energy or quasienergy (QE) band spectra depending on two parameters may have a nontrivial topological characterization by Chern integers. Band spectra of 1D systems that are spanned by just one parameter, a Bloch phase, are topologically trivial. Recently, an ensemble of 1D Floquet (time-periodic) systems, double kicked rotors (DKRs) that are classically nonintegrable and depend on an external parameter, has been studied. It was shown that a QE band spanned by both the Bloch phase and the external parameter is characterized by a Chern integer. The latter determines the change in the mean angular momentum of a state in the band when the external parameter is adiabatically varied by a natural period. We show here, under conditions much more general than in previous works, that the ensemble of DKRs for all values of the external parameter corresponds to a 1D double kicked particle (DKP) having translational invariance in the position-momentum phase plane. This DKP can be characterized by a \emph{second} Chern integer which is shown to be connected with the integer above for the DKR ensemble. This connection is expressed by a Diophantine equation (DE) which we derive. The DE, involving the band degeneracies of the DKR ensemble and of the DKP system, determines the allowed values of the DKR-ensemble integer. In particular, this integer is generically nonzero, showing the general topological nontriviality of the DKR ensemble. 
\end{abstract}

\maketitle

\begin{center}
\textbf{I. INTRODUCTION}
\end{center}

Band spectra depending on two parameters, such as Bloch phases, may be topologically characterized by nonzero Chern integers \cite{tknn,ahm,k,daz,dz,id00,df,dc,id0,qc1,qc2,qc3,qc4,id,id1} having some physical significance, e.g., in the context of quantum transport. A well known example is the system of 2D Bloch electrons in a uniform magnetic field \cite{tknn,ahm,k,daz,dz,id00}. As shown in the pioneering paper by Thouless, Kohmoto, Nightingale, and den Nijs (TKNN) \cite{tknn}, the quantum Hall conductance of this system in linear-response theory is given by $e^2/h$ times the total Chern integer of the magnetic bands below the Fermi level. Following works on this static (time-independent) system, a first study of the topological characterization of Floquet (time-periodic) systems was performed by Leboeuf {\it et al.} \cite{qc1} for the ``kicked Harper" model (KHM) \cite{qc1,qc2,qc3,qc4,id,id1}; a generalized version of this system is given by the Hamiltonian \cite{id,id1}  
\begin{equation}\label{KHM}
\hat{H}_{\mathrm{KHM}}=W_p(\hat{p})+W_x(\hat{x})\sum_{s=-\infty}^{\infty}\delta(t-s)
\end{equation}
in the position-momentum $(\hat{x},\hat{p})$ phase plane, where $W_p$ and $W_x$ are $2\pi$-periodic functions \cite{note} and we choose units such that the time period is unity. The KHM is a classically nonintegrable system, exhibiting chaos, and is realistic, being exactly related to periodically kicked charges in a magnetic field \cite{id}. For rational values of an effective Planck constant $\hbar_{\rm eff}$, the quasienergy (QE) spectrum of (\ref{KHM}) consists of bands characterized by Chern integers \cite{id1}. In a semiclassical regime, zero or nonzero values of this integer are usually associated with QE eigenstates localized on classical regular orbits or spread over the chaotic region, respectively \cite{qc1,qc2,qc3}. 

An important feature of the static TKNN system and the Floquet KHM (\ref{KHM}) is translational invariance on some phase plane, e.g., $(\hat{x},\hat{p})$ for the KHM. This invariance implies that a band can be characterized by \emph{two} Chern integers which must satisfy a Diophantine equation (DE) for a rational value of some parameter. The latter is the number of flux quanta per unit cell in the TKNN case \cite{tknn,ahm,daz,dz} or $\hbar_{\rm eff}$ in the KHM case \cite{id1}. The DE has many applications \cite{tknn,ahm,daz,dz,id00,id0}, e.g., it unambiguously determines the Chern integer of a band in several systems \cite{tknn,ahm} as well as the total Chern integer of general spectral clusters (including the full cluster up to a Fermi level) in limits of irrational number of flux quanta per unit cell in the TKNN case \cite{daz,id0}.

More recently, there has been a considerable new interest in the topological properties of Floquet systems \cite{fti,tcfs,lsz,jg1,jg2,jg3}, especially in the context of in-gap edge states and Floquet topological insulators. A class of systems that has been studied on a first-principles basis are the nonintegrable double kicked rotors (DKRs) \cite{jg1,jg2,jg3,jg4,id2}, defined by the general Hamiltonian \cite{id2}
\begin{eqnarray}
\hat{H}_{\mathrm{DKR}} &=&\frac{\hat{L}^{2}}{2}+KV(\theta )\sum_{s=-\infty }^{\infty
}\delta (t-s)  \notag \\
&+&\tilde{K}\tilde{V}(\theta )\sum_{s=-\infty }^{\infty }\delta (t-\eta -s),  \label{HDKR}
\end{eqnarray}
where $(\theta ,\hat{L})$ are angle and angular momentum, $K$ and $\tilde{K}$ are nonintegrability parameters, $V(\theta )$ and $\tilde{V}(\theta )$ are
general $2\pi$-periodic potentials, and $\eta$ is the second kicking time ($0\leq \eta <1$). For rational values of $\eta$ and of some $\hbar_{\rm eff}$, the QE spectrum of (\ref{HDKR}) consists of bands spanned by just one Bloch phase \cite{id2}. Such bands are therefore topologically trivial. Recent works \cite{jg1,jg2,jg3} have considered the case of cosine potentials in Eq. (\ref{HDKR}), with one potential depending on an external parameter $\alpha$, under some ``on-resonance" (main quantum-resonance \cite{id2}) conditions. When a QE band is viewed as depending on both the Bloch phase and $\alpha$, it may turn out to be topologically nontrivial, featuring a nonzero Chern integer \cite{jg1,jg2}. This integer \emph{exactly} determines the change in the mean angular momentum of a state in the band when $\alpha$ is adiabatically varied by $2\pi$ \cite{jg1}. It was also shown \cite{jg2} that the ensemble of on-resonance DKRs (ODKRs) for all $\alpha$ is unitarily equivalent to a a similar ensemble of kicked Harper rotors (KHRs), i.e., Eq. (\ref{KHM}) with $(\hat{x},\hat{p})\rightarrow (\theta ,\hat{L})$. A KHR is topologically trivial like a single DKR. 

In this paper, we investigate topological properties of ODKRs depending on an external parameter $\alpha$ under conditions much more general than in previous works \cite{jg1,jg2,jg3}: We consider arbitrary $2\pi$-periodic potentials in Eq. (\ref{HDKR}) and general on-resonance conditions. We show that the ensemble of the generalized ODKRs for all $\alpha$ corresponds to a 1D double-kicked-particle (DKP) system possessing translational invariance in the $(\hat{x},\hat{p})$ phase plane. This system is essentially a generalized KHM. A ODKR for some parameter $\alpha$ is given by the restriction of the DKP system to a fixed value of the quasimomentum. This result is similar to that connecting ODKRs with KHRs in Ref. \cite{jg2} (see above) but its derivation is much simpler and it holds under the general conditions above. More important, the DKP system is characterized by a \emph{second} Chern integer, which is shown to be connected with the ODKR-ensemble integer characterizing the adiabatic quantum transport in Ref. \cite{jg1} (see also above). The connection between the two Chern integers is expressed by a DE which we derive. This DE, which involves the degeneracy of the QE band eigenstates of the ODKR ensemble and of the DKP system, determines the allowed values of the ODKR-ensemble integer. In particular, this integer is generically nonzero, showing the general topological nontriviality of the ODKR ensemble.

The paper is organized as follows. In Sec. II, we present a general background containing a summary of previous results in terms of the generalized DKRs (\ref{HDKR}). In Sec. III, we introduce our DKP system, having translational invariance in the $(\hat{x},\hat{p})$ phase plane, and establish the correspondence between it and the generalized ODKR ensemble. In Sec. IV, we study basic properties of the QE band spectrum and eigenstates of the DKP system and derive an explicit general expression for the eigenstates. This is then related, using the correspondence above, to the QE band spectrum and eigenstates of the ODKR ensemble. In Sec. V, we study topological and related properties of the QE band eigenstates of the DKP system and of the ODKR ensemble. Expressions for the two topological Chern integers associated with these systems are derived. The adiabatic quantum transport in Ref. \cite{jg1}, characterized by the ODKR-ensemble integer, is then discussed using our results. Our main result, a DE connecting the two Chern integers, is derived in Sec. VI. In Sec. VII, we provide specific examples to illustrate some concepts and to verify the validity of the DE. A summary and conclusions are presented in Sec. VIII.

\begin{center}
\textbf{II. GENERAL BACKGROUND}
\end{center} 

In order to provide a basis and motivation for the study in the next sections, we summarize here relevant previous results \cite{jg1,jg2}, but in a more general framework. Consider the general Hamiltonian (\ref{HDKR}) with the potential $\tilde{V}(\theta )$ replaced by $\tilde{V}(\theta +\alpha )$, where $\alpha$ is a parameter. The corresponding one-period evolution operator, from $t=s-0$ to $t=s+1-0$, is given by
\begin{eqnarray}
\hat{U}_{\rm DKR}^{(\alpha )}(\theta,\hat{L}) &=&\exp \left[ -i\left( 1-\eta \right)\frac{\hat{L}
^{2}}{2\hbar }\right] \exp \left[ -i\frac{\tilde{K}\tilde{V}(\theta +\alpha )}{\hbar }\right]  \notag \\
&\times &\exp \left( -i\eta \frac{\hat{L}^{2}}{2\hbar }\right) \exp \left[ -i\frac{KV(\theta )}{\hbar }\right] .  \label{UDKR}
\end{eqnarray} 
where $\hbar$ is a scaled (dimensionless) Planck constant in our units. We assume the general on-resonance conditions (corresponding to main quantum resonances, see Ref. \cite{id2} and below):
\begin{equation}\label{orc}
\hbar =2\pi l ,
\end{equation}
where $l$ is an arbitrary \emph{even} integer. Previous works \cite{jg1,jg2,jg3,jg4} were restricted to the special case of cosine potentials in Eq. (\ref{UDKR}) and $l=2$ in Eq. (\ref{orc}). In general, under condition (\ref{orc}), the operator (\ref{UDKR}) reduces to the ODKR evolution operator (see details in Appendix A) 
\begin{eqnarray}
\hat{U}_{\rm ODKR}^{(\alpha )}(\theta,\hat{L}) &=&\exp \left( i\eta \frac{\hat{L}
^{2}}{2\hbar }\right) \exp \left[ -i\frac{\tilde{K}\tilde{V}(\theta +\alpha)}{\hbar }\right]  \notag \\
&\times &\exp \left( -i\eta \frac{\hat{L}^{2}}{2\hbar }\right) \exp \left[ -i\frac{KV(\theta )}{\hbar }\right] .  \label{UODKR}
\end{eqnarray} 
For arbitrary $\hbar$, not necessarily satisfying the condition (\ref{orc}), the operator (\ref{UODKR}) corresponds to the Hamiltonian of a modified kicked-rotor model studied in work \cite{gw}.

We now require the operator (\ref{UODKR}) to be invariant under translations in angular momentum $\hat{L}$ by $N\hbar$, for some integer $N$. This implies that $\eta$ must be rational (see details in Appendix A),
\begin{equation}\label{eta}
\eta =\frac{a}{c},
\end{equation}
where $a$ and $c$ are coprime integers. Given $\eta$ by Eq. (\ref{eta}) and writing $l/c=l'/c'$, where $l'$ and $c'$ are coprime integers, the minimal value of $N$ is (see Appendix A):
\begin{equation}\label{N}
N =c' .
\end{equation}  

The translational invariance of the operator (\ref{UODKR}) in $\hat{L}$ by $N\hbar=c'\hbar$ implies that its eigenstates, i.e., the QE states, have a Bloch form in the $\hat{L}$ representation: 
\begin{equation}\label{qes}
\Phi_{b,\gamma}^{(\alpha )}(n)=\exp(-i\gamma n)\phi_{b,\gamma}^{(\alpha )}(n),\ \ \phi_{b,\gamma}^{(\alpha )}(n+c')=\phi_{b,\gamma}^{(\alpha )}(n), 
\end{equation}
where the index $b$ labels QE bands $\omega_b^{(\alpha )}(\gamma )$, each spanned by the Bloch phase $\gamma$. These bands define the eigenvalues $\exp [-i\omega_b^{(\alpha )}(\gamma )]$ of the operator (\ref{UODKR}), corresponding to the eigenstates $\Phi_{b,\gamma}^{(\alpha )}(n)$. As shown in Sec. IV, the QE spectrum consists of precisely $c'$ bands, $b=1,...,c'$.

In general, a QE band continuous spectrum of the operator (\ref{UDKR}) at fixed $\alpha$ occurs for rational values of both $\eta$ and $\hbar/(2\pi )$ \cite{id2} and leads to quantum resonance, i.e., the expectation value of the kinetic energy $\hat{L}^2/2$ increases quadratically in time [number of applications of the operator (\ref{UDKR})]. The values (\ref{orc}) of $\hbar$ correspond to the main (strongest) quantum resonances.

At fixed $\alpha$, however, each band is spanned by a single Bloch phase, $\gamma$, and is therefore topologically trivial. On the other hand, when a band is viewed as depending on both $\alpha$ and $\gamma$, it may be topologically nontrivial, characterized by a nonzero Chern integer \cite{jg1,jg2}, see also Secs. V and VI. This integer exactly determines the change in the mean angular momentum of a state in a band when $\alpha$ is adiabatically varied by $2\pi$ \cite{jg1}. 

\begin{center}
\textbf{III. CORRESPONDENCE BETWEEN THE ODKR ENSEMBLE AND A DKP SYSTEM}
\end{center}           

To derive our main results concerning topological and related properties of the generalized ensemble of ODKR operators (\ref{UODKR}) for all $\alpha$, under the conditions (\ref{orc}) and (\ref{eta}), we proceed in four stages: (a) In this section, we establish a useful correspondence between the ODKR ensemble and a double-kicked-particle (DKP) system. (b) In Sec. IV, we study basic properties of the QE band spectrum and eigenstates of the DKP system and derive an explicit general expression for the eigenstates. This is then related, using the correspondence above, to the QE band spectrum and eigenstates of the ODKR ensemble. (c) In Sec. V, we write the periodicity conditions satisfied by the QE eigenstates of the DKP system and of the ODKR ensemble. This is then used to derive expressions for two topological Chern integers, one associated with the DKP system and the other with the ODKR ensemble. (d) Using the latter expressions, we derive in Sec. VI a DE connecting the two Chern integers.

We define the DKP system by its evolution operator, given by Eq. (\ref{UODKR}) for $\alpha =0$ with $(\theta,\hat{L})$ replaced by ordinary position and momentum $(\hat{x},\hat{p})$ on a phase plane:
\begin{eqnarray}
\hat{U}_{\rm DKP}(\hat{x},\hat{p}) &=&\exp \left( i\eta \frac{\hat{p}
^{2}}{2\hbar }\right) \exp \left[ -i\frac{\tilde{K}\tilde{V}(\hat{x} )}{\hbar }\right]  \notag \\
&\times &\exp \left( -i\eta \frac{\hat{p}^{2}}{2\hbar }\right) \exp \left[ -i\frac{KV(\hat{x} )}{\hbar }\right] .  \label{UP}
\end{eqnarray}
Actually, the DKP operator is given by Eq. (\ref{UDKR}) for $\alpha =0$ with $(\theta,\hat{L})$ replaced by $(\hat{x},\hat{p})$. The operator (\ref{UP}) may be viewed as corresponding to a modified DKP, analogous to the modified kicked rotor studied in work \cite{gw}. Using the identity $e^{i\eta\hat{p}^2/(2\hbar )}f(\hat{x})e^{-i\eta\hat{p}^2/(2\hbar )}=f(\hat{x}+\eta\hat{p})$ for arbitrary function $f$ (see also Ref. \cite{jg4}), the operator (\ref{UP}) can be expressed as
\begin{equation}\label{UPn}
\hat{U}_{\rm DKP}(\hat{x},\hat{p}) =\exp \left[ -i\frac{\tilde{K}\tilde{V}(\hat{x}+\eta\hat{p} )}{\hbar }\right] \exp \left[ -i\frac{KV(\hat{x} )}{\hbar }\right] .
\end{equation}
Equation (\ref{UPn}) gives precisely the evolution operator of a generalized KHM (\ref{KHM}) with $\hat{p}$ replaced by a new momentum $\hat{p}'=\hat{x}+\eta\hat{p}$. Using $[\hat{x},\hat{p}']=i\hbar '=i\eta\hbar$, some results below follow straightforwardly from similar results for the KHM in Ref. \cite{id1}. However, in order to establish a relation between the operator (\ref{UPn}) and the ODKR one as clearly and simply as possible, we shall not transform to the momentum $\hat{p}'$. Also, for the convenience of the reader, our presentation will be self-contained and more detailed than in Ref. \cite{id1}.

To derive a relation [Eq. (\ref{eqr}) below] between the operator (\ref{UP}) [or (\ref{UPn})] and the ODKR one (\ref{UODKR}), we first note that the operator (\ref{UPn}) clearly commutes with a $2\pi$-translation in $\hat{x}$:
\begin{equation}\label{Dx}
\hat{D}_{x,2\pi}=\exp(2\pi i\hat{p}/\hbar )=\exp(2\pi d/dx),
\end{equation}
where we used $\hat{p}=-i\hbar d/dx$ (from $[\hat{x},\hat{p}]=i\hbar$). The simultaneous eigenstates of (\ref{UPn}) and (\ref{Dx}) can then be chosen to have the Bloch form in the $x$ representation:
\begin{equation}\label{bf}
\left\langle x\right|\left.\Psi_{\beta}\right\rangle = \exp (i\beta x)\psi_{\beta}(x),\ \ \psi_{\beta}(x+2\pi )=\psi_{\beta}(x),
\end{equation}
where $\beta$, $0\leq\beta <1$, gives the quasimomentum $\beta\hbar$ and specifies the eigenvalues of both the operator (\ref{UPn}),
\begin{equation}\label{eUP}
\left\langle x\right|\hat{U}_{\rm DKP}(\hat{x},\hat{p})\left|\Psi_{\beta}\right\rangle=\exp [-i\omega (\beta)]\left\langle x\right|\left.\Psi_{\beta}\right\rangle,
\end{equation}
and the operator (\ref{Dx}),
\begin{equation}\label{eDx}
\hat{D}_{x,2\pi}\left|\Psi_{\beta}\right\rangle = \exp (2\pi i\beta )\left|\Psi_{\beta}\right\rangle,
\end{equation}
where $\omega (\beta)$ in Eq. (\ref{eUP}) are the QEs. Inserting Eqs. (\ref{UPn}) and (\ref{bf}) into Eq. (\ref{eUP}) and  multiplying both sides of Eq. (\ref{eUP}) by $\exp (-i\beta x)$, which is a translation of $\hat{p}$ by $\beta\hbar$ (since $\hat{x}=i\hbar d/dp$), we find that $\psi_{\beta}(x)$ is an eigenstate of an operator $\hat{U}_{\rm DKP}(\hat{x},\hat{p}+\beta\hbar )$ [Eq. (\ref{UPn}) with $\hat{p}$ replaced by $\hat{p}+\beta\hbar$]. Then, in the same way Eq. (\ref{UPn}) can be expressed as Eq. (\ref{UP}), $\hat{U}_{\rm DKP}(\hat{x},\hat{p}+\beta\hbar )$ can be expressed as 
\begin{eqnarray}\label{UPb}
\hat{U}_{\rm DKP}(\hat{x},\hat{p}+\beta\hbar ) &=&\exp \left( i\eta \frac{\hat{p}
^{2}}{2\hbar }\right) \exp \left[ -i\frac{\tilde{K}\tilde{V}(\hat{x}+\beta\eta\hbar )}{\hbar }\right]  \notag \\
&\times &\exp \left( -i\eta \frac{\hat{p}^{2}}{2\hbar }\right) \exp \left[ -i\frac{KV(\hat{x} )}{\hbar }\right] .
\end{eqnarray}
Now, since the eigenstates $\psi_{\beta}(x)$ of the operator (\ref{UPb}) are $2\pi$-periodic in $x$ [see Eq. (\ref{bf})], one can interpret $(\hat{x},\hat{p})$ in Eq. (\ref{UPb}) as angle and angular momentum $(\theta ,\hat{L})$. Then, by comparing Eq. (\ref{UPb}) with Eq. (\ref{UODKR}), we see that
\begin{equation}\label{eqr}
\hat{U}_{\rm DKP}(\theta ,\hat{L}+\beta\hbar )=\hat{U}_{\rm ODKR}^{(\alpha)}(\theta,\hat{L}), \ \ \ \alpha=\beta\eta\hbar .
\end{equation}
Equation (\ref{eqr}) establishes the correspondence between the DKP operator (\ref{UPb}) at fixed $\beta$ and the ODKR operator (\ref{UODKR}) with $\alpha =\beta\eta\hbar$. We shall use this correspondence to re-derive and significantly extend results in previous works, mentioned in Sec. II.

\begin{center}
\textbf{IV. QE SPECTRUM AND EIGENSTATES}
\end{center}

Relation (\ref{eqr}) was derived just on the basis of the invariance of the operator (\ref{UPn}) under $2\pi$-translations (\ref{Dx}) in $\hat{x}$. To get more information about the QE spectrum and eigenstates, we note that the operator (\ref{UPn}) clearly commutes also with a minimal translation of $\hat{p}$ by $2\pi/\eta$:
\begin{equation}\label{Dp}
\hat{D}_{p,2\pi/\eta}=\exp[(2\pi/\eta )d/dp]=\exp[-2\pi i\hat{x}/(\eta\hbar )].
\end{equation}      
For general value (\ref{eta}) of $\eta$, the phase-space translations (\ref{Dx}) and (\ref{Dp}) do not commute:
\begin{equation}\label{Dxp}
\hat{D}_{x,2\pi}\hat{D}_{p,2\pi/\eta}=\exp[-4\pi ^2 i/(\eta\hbar )] \hat{D}_{p,2\pi/\eta}\hat{D}_{x,2\pi} .
\end{equation}
However, from Eq. (\ref{orc}), Eq. (\ref{eta}), and the definition above of the coprime integers $(l',c')$ by $l/c=l'/c'$, we see that $4\pi^2/(\eta\hbar )=2\pi c'/(al' )$. Then, Eq. (\ref{Dxp}) implies that the minimal translation in $\hat{p}$ commuting with both operators (\ref{UPn}) and (\ref{Dx}) is 
\begin{equation}\label{Dpc}
\left(\hat{D}_{p,2\pi/\eta}\right)^{al'}=\exp(c'\hbar d/dp)=\exp(-ic'\hat{x}).
\end{equation} 
This is a translation of $\hat{p}$ by $c'\hbar$, in consistency with Eq. (\ref{N}). Since the operators (\ref{UPn}), (\ref{Dx}), and (\ref{Dpc}) commute with each other, the QE eigenstates $\left|\Psi_{\beta}\right\rangle$ in Eqs. (\ref{eUP}) and (\ref{eDx}) are also eigenstates of the operator (\ref{Dpc}):
\begin{equation}\label{eDpc}
\exp(-ic'\hat{x})\left|\Psi_{\beta}\right\rangle = \exp (-ic'\gamma ) \left|\Psi_{\beta}\right\rangle  .
\end{equation} 
As shown below, at fixed values of $\beta$ and $\gamma$ the QE spectrum consists of precisely $c'$ levels $\omega_b(\beta ,\gamma )$, $b=1,...,c'$. We shall assume from now on that the corresponding QE eigenvalues are non-degenerate, i.e., $\exp[-i\omega_b(\beta ,\gamma )]\neq \exp[-i\omega_{b'}(\beta ,\gamma )]$ for $b\neq b'$ and for all $(\beta ,\gamma )$. Then, as $\beta$ and $\gamma$ cover the ranges of the eigenvalues in Eqs. (\ref{eDx}) and (\ref{eDpc}), i.e., the Brillouin zone (BZ)
\begin{equation}\label{BZ}
0\leq\beta <1,\ \ \ 0\leq\gamma <2\pi/c' ,
\end{equation}
each level $b$ will broaden into a QE band $\omega_b(\beta ,\gamma )$ and these bands are isolated (do not cross). Due to the equivalence relation (\ref{eqr}) between the DKP system and the ODKR ensemble, one has:
\begin{equation}\label{eqrqe}
\omega_b(\beta,\gamma )=\omega_b^{(\alpha )}(\gamma ) ,\ \ \alpha = \beta\eta\hbar.
\end{equation}

We now derive a general expression for the QE eigenstates $\left|\Psi_{\beta}\right\rangle =\left|\Psi_{b,\beta,\gamma}\right\rangle$ associated with the bands $\omega_b(\beta,\gamma )$. First, Eqs. (\ref{Dpc}) and (\ref{eDpc}) imply that the momentum ($p$) representation of $\left|\Psi_{b,\beta,\gamma}\right\rangle$ must have the Bloch form
\begin{equation}\label{bfp}
\left\langle p\right|\left.\Psi_{b,\beta,\gamma}\right\rangle=\exp (-i\gamma p/\hbar)\lambda_{b,\beta,\gamma}(p),
\end{equation}
where $\lambda_{b,\beta,\gamma}(p+c'\hbar)=\lambda_{b,\beta,\gamma}(p)$. Second, it is clear from Eq. (\ref{bf}) that $\left\langle p\right|\left.\Psi_{b,\beta,\gamma}\right\rangle$ can exhibit only momenta $(n+\beta )\hbar$ for all integers $n$. Thus, $\lambda_{b,\beta,\gamma}(p)$ in Eq. (\ref{bfp}) must be a linear combination of delta functions $\delta(p-n\hbar-\beta\hbar )$ periodic in $n$ with period $c'$. Then, after setting $p=(n+\beta )\hbar$ in the phase factor $\exp (-i\gamma p/\hbar )$ in Eq. (\ref{bfp}) and omitting a constant factor $\exp (-i\beta\gamma )$, we obtain the following expression for the QE eigenstates:
\begin{eqnarray}\label{qesp}
\left\langle p\right|\left.\Psi_{b,\beta,\gamma}\right\rangle & = & \sum_{n=0}^{c'-1}\Phi_{b,\beta,\gamma}(n) \\ \notag 
& \times & \sum_{j=-\infty}^{\infty}e^{-ijc'\gamma }\delta(p-n\hbar-\beta\hbar-jc'\hbar ). 
\end{eqnarray}
Here the coefficients $\Phi_{b,\beta,\gamma}(n)$ include a phase factor $\exp (-i\gamma n)$. Now, one has the equivalence relation
\begin{equation}\label{eqrc}
\Phi_{b,\beta,\gamma}(n)= \Phi_{b,\gamma}^{(\alpha )}(n),\ \ \alpha = \beta\eta\hbar , 
\end{equation} 
where $\Phi_{b,\gamma}^{(\alpha )}(n)$ are the states (\ref{qes}). To show Eq. (\ref{eqrc}), we note from the derivation of Eq. (\ref{qesp}) that the $p$ representation $\chi_{b,\beta,\gamma}(p)$ of the periodic function $\psi_{\beta}(x)$ in Eq. (\ref{bf}) is given by Eq. (\ref{qesp}) with $\beta =0$ in the delta functions:
\begin{eqnarray}\label{qespp}
\chi_{b,\beta,\gamma}(p) & = & \sum_{n=0}^{c'-1}\Phi_{b,\beta,\gamma}(n) \\ \notag 
& \times & \sum_{j=-\infty}^{\infty}e^{-ijc'\gamma }\delta(p-n\hbar-jc'\hbar ). 
\end{eqnarray} 
On the other hand, from Eq. (\ref{eqr}) and the fact that $\psi_{\beta}(x)$ is an eigenstate of the operator (\ref{UPb}) it is clear that $\Phi_{b,\beta,\gamma}(n)$ in Eq. (\ref{qespp}) must be equal to the eigenstate (\ref{qes}) of the ODKR operator (\ref{UODKR}); the Bloch phase factors $\exp (-ijc'\gamma )$ just express the quasi-periodicity of (\ref{qes}) in $n$ with period $N=c'$. This shows Eq. (\ref{eqrc}). Equations (\ref{eqr}), (\ref{eqrqe}), and (\ref{eqrc}) fully establish the correspondence between the DKP system and the ODKR ensemble.

Since the QE eigenstates (\ref{qesp}) must form a complete set of states and there can only be $c'$ independent sets of coefficients $\{ \Phi_{b,\beta,\gamma}(n)\}_{n=0}^{c'-1}$, the number of QE bands must be exactly $c'$, $b=1,...,c'$, as mentioned in Sec. II. 

\begin{center}
\textbf{V. TOPOLOGICAL PROPERTIES, ADIABATIC QUANTUM TRANSPORT, AND DEGENERACIES}
\end{center}

In this section, we study topological and related properties of the QE band eigenstates. After writing periodicity conditions for these eigenstates in suitable Brillouin zones, we derive general expressions for two topological Chern integers, one ($S_b$) associated with the eigenstates of the DKP system and the other ($C_b$) with the eigenstates of the ODKR ensemble. The integer $C_b$ is shown to be that characterizing the adiabatic quantum transport of ODKR states in Ref. \cite{jg1}. 

From our assumption of isolated (non-crossing) bands, see previous section, it follows that the QE eigenstates $\left|\Psi_{b,\beta,\gamma}\right\rangle$  must be periodic in the BZ (\ref{BZ}) up to constant phase factors: 
\begin{eqnarray}
\left|\Psi_{b,\beta+1,\gamma}\right\rangle & = & \exp [if_b(\beta ,\gamma )]\left|\Psi_{b,\beta,\gamma}\right\rangle , \label{pqe1} \\ 
\left|\Psi_{b,\beta,\gamma +2\pi /c'}\right\rangle & = & \exp [ig_b(\beta ,\gamma )]\left|\Psi_{b,\beta,\gamma}\right\rangle . \label{pqe2}  
\end{eqnarray} 
We now determine the periodicity in $(\beta ,\gamma )$ of $\Phi_{b,\beta,\gamma}(n)$ in Eq. (\ref{qesp}); $\Phi_{b,\beta,\gamma}(n)$ give the ODKR eigenstates by Eq. (\ref{eqrc}). Let us apply the translation (\ref{Dp}) to Eq. (\ref{qesp}):
\begin{eqnarray}\label{dqesp}
\left\langle p\right|\hat{D}_{p,2\pi/\eta}\left|\Psi_{b,\beta,\gamma}\right\rangle & = & \sum_{n=0}^{c'-1}\Phi_{b,\beta,\gamma}(n)\sum_{j=-\infty}^{\infty}e^{-ijc'\gamma } \\ \notag 
& \times & \delta\left( p-n\hbar-\beta\hbar+\frac{2\pi}{\eta}-jc'\hbar \right). 
\end{eqnarray}
From Eqs. (\ref{Dxp}) and (\ref{eDpc}), it follows that the state (\ref{dqesp}) satisfies Eq. (\ref{eDx}) with $\beta$ replaced by $\beta -2\pi/(\eta\hbar )=\beta -c'/(al')$. Moreover, since $\hat{D}_{p,2\pi/\eta}$ commutes with the operator (\ref{UPn}), the state (\ref{dqesp}) is degenerate with $\left|\Psi_{b,\beta,\gamma}\right\rangle$ and thus belongs to band $b$ because of the assumption of isolated bands. Therefore, the state (\ref{dqesp}) must be equal to $\left|\Psi_{b,\beta -c'/(al'),\gamma}\right\rangle$ up to a constant phase factor:
\begin{equation}\label{dqespe}
\hat{D}_{p,2\pi/\eta}\left|\Psi_{b,\beta,\gamma}\right\rangle = \exp [iu_b(\beta ,\gamma )]\left|\Psi_{b,\beta -c'/(al'),\gamma}\right\rangle .
\end{equation}
Equations (\ref{qesp}), (\ref{dqesp}), and (\ref{dqespe}) imply that $\Phi_{b,\beta -c'/(al'),\gamma}(n)$ is equal to $\Phi_{b,\beta,\gamma}(n)$ up to the phase factor in Eq. (\ref{dqespe}). Equation (\ref{qesp}) also implies that $\Phi_{b,\beta,\gamma +2\pi /c'}(n)$ is equal to $\Phi_{b,\beta,\gamma}(n)$ up to the phase factor in Eq. (\ref{pqe2}). In summary:
\begin{eqnarray}
\Phi_{b,\beta +c'/(al'),\gamma}(n) & = & \exp [iw_b(\beta ,\gamma )]\Phi_{b,\beta ,\gamma}(n) , \label{pqep1} \\ 
\Phi_{b,\beta,\gamma +2\pi /c'}(n) & = & \exp [ig_b(\beta ,\gamma )]\Phi_{b,\beta ,\gamma}(n) , \label{pqep2}   
\end{eqnarray}
where $w_b(\beta ,\gamma )=u_b(\beta +c'/(al') ,\gamma )$.

Due to the single-valuedness of $\left|\Psi_{b,\beta,\gamma}\right\rangle$ in the BZ (\ref{BZ}), the total phase change of $\left|\Psi_{b,\beta,\gamma}\right\rangle$ when going around the BZ boundary counterclockwise must be equal to an integer multiple $S_b$ of $2\pi$. The quantity $S_b$ is a topological Chern integer for band $b$, see, e.g., works
\cite{tknn,ahm,k,daz,dz,id00,df,dc,id0,qc1,qc2,qc3,qc4,id,id1}. Let us derive an expression for $S_b$. From Eqs. (\ref{pqe1}) and (\ref{pqe2}), it follows that when $\gamma$ is varied from $\gamma$ to $\gamma +2\pi /c'$ (on the vertical axis of the BZ) the total change in the phase of $\left|\Psi_{b,\beta +1,\gamma }\right\rangle$ relative to that of $\left|\Psi_{b,\beta,\gamma}\right\rangle$ is $f_b(\beta, \gamma +2\pi /c')-f_b(\beta, \gamma )$; when $\beta$ is varied from $\beta +1$ to $\beta$ (on the horizontal axis of the BZ) the total change in the phase of $\left|\Psi_{b,\beta,\gamma +2\pi /c'}\right\rangle$ relative to that of $\left|\Psi_{b,\beta ,\gamma}\right\rangle$ is $g_b(\beta, \gamma )-g_b(\beta +1, \gamma )$. Thus, one must have
\begin{eqnarray}\label{ci1}
f_b\left(\beta, \gamma +\frac{2\pi}{c'}\right) & - & f_b(\beta, \gamma ) \\ \notag & + & g_b(\beta, \gamma ) -g_b(\beta +1, \gamma ) =2\pi S_b . 
\end{eqnarray}
 
Similarly, from Eqs. (\ref{pqep1}) and (\ref{pqep2}) one can associate another Chern integer $C_b$ with band $b$:           
\begin{eqnarray}\label{ci2}
w_b\left(\beta, \gamma +\frac{2\pi}{c'}\right) & - & w_b(\beta, \gamma ) \\ \nonumber & + & g_b(\beta, \gamma )-g_b\left(\beta +\frac{c'}{al'}, \gamma \right) =2\pi C_b . 
\end{eqnarray}
The integer $C_b$ is precisely the ODKR-ensemble Chern integer determining, as in work \cite{jg1} (where only cosine potentials were considered), the change in the mean angular momentum of an ODKR state in band $b$ when the parameter $\alpha$ in Eq. (\ref{UODKR}) is adiabatically varied from $0$ to $2\pi$. To see this, we note that the shift of $\beta$ by $c'/(al')=2\pi /(\eta\hbar )$ in Eq. (\ref{pqep1}) corresponds, by Eq. (\ref{eqrc}), to a shift of $\alpha$ by $2\pi$. Then, representing an ODKR eigenstate by the column vector ${\mathbf V}_b({\mathbf w})=\{\Phi_{b,\gamma }^{(\alpha )}(n)\}_{n=0}^{c'-1}$, where ${\mathbf w}=(\alpha ,\gamma )$, and using Eq. (\ref{eqrc}), we can rewrite Eq. (\ref{ci2}) as 
\begin{equation}\label{Cb}
C_b=\frac{1}{2\pi i}\varoint_{\rm BZ1}{\mathbf V}_b^{\dagger}({\mathbf w})\frac{d{\mathbf V}_b({\mathbf w})}{d{\mathbf w}}\cdot d{\mathbf w}. 
\end{equation}
Here BZ1 is the Brillouin zone for the ODKR eigenstates,
\begin{equation}\label{BZ1}
0\leq \alpha <2\pi ,\ \ \ 0\leq \gamma <2\pi /c' ,
\end{equation}
and the contour integral is performed counterclockwise around the boundary of the zone (\ref{BZ1}). Using Stoke's theorem, the latter integral can be expressed as the integral of a Berry's curvature over BZ1. The resulting formula for $C_b$ can be easily seen to coincide with that given for cosine potentials by Eq. (9) in Ref. \cite{jg1} [or Eq. (12) in Ref. \cite{jg2}], but it now holds for arbitrary potentials.

We remark that $C_b$ gives the total vorticity of all the zeros \cite{k,df,dc} of $\Phi_{b,\gamma}^{(\alpha )}(n)$ in the zone (\ref{BZ1}). Similarly, $S_b$ is the total vorticity of all the zeros of the QE eigenstates $\left|\Psi_{b,\beta,\gamma}\right\rangle$ in the BZ (\ref{BZ}), when these eigenstates are expressed in a suitable representation, e.g., the coherent-state one as in Ref. \cite{qc3}, rather than the $p$ representation of Eq. (\ref{qesp}). While we do not have yet a quantum-transport interpretation of $S_b$, this integer determines the allowed values of $C_b$, as we show in the next section. 

As mentioned above, the state (\ref{dqesp}) is degenerate with $\left|\Psi_{b,\beta,\gamma}\right\rangle$ and is associated with the quasimomentum $\beta -c'/(al')$, see Eq. (\ref{dqespe}). In general, the $al'$ states $\hat{D}_{p,2\pi/\eta}^s\left|\Psi_{b,\beta,\gamma}\right\rangle$, $s=0,1,...,al'-1$, are all degenerate eigenstates in band $b$, associated with the quasimomenta $\beta -sc'/(al')\ {\rm mod}(1)$. Thus, a QE band $\omega_b(\beta,\gamma )$ must be periodic in $\beta$ with period $c'/(al')$ and also with period $1$ [in the BZ (\ref{BZ})]. This is possible only if the minimal periodicity zone of $\omega_b(\beta,\gamma )$ is       
\begin{equation}\label{QEBZ}
0\leq \beta <1/(al'),\ \ \ 0\leq \gamma <2\pi /c' .
\end{equation}
Using Eq. (\ref{eqrc}), the corresponding periodicity zone for the ODKR QE bands $\omega_b^{(\alpha )}(\gamma )$ is 
\begin{equation}\label{QEBZ1}
0\leq \alpha <2\pi /c',\ \ \ 0\leq \gamma <2\pi /c' .
\end{equation}
This means that the $c'$ ODKR eigenstates $\Phi_{b,\gamma }^{(\alpha +2\pi s/c') }(n)$, $s=0,1,...,c'-1$, are all degenerate; see also note \cite{note1}. 

\begin{center}
\textbf{VI. DIOPHANTINE EQUATION (DE) FOR CHERN INTEGERS}
\end{center}

We derive here a DE for the Chern integers $S_b$ and $C_b$. Let us first iterate Eq. (\ref{pqe1}) $c'$ times; this gives
\begin{equation}\label{dcqespe}
\left |\Psi_{b,\beta +c',\gamma}\right\rangle = \exp [i\bar{f}_b(\beta ,\gamma )]\left |\Psi_{b,\beta ,\gamma}\right\rangle ,
\end{equation}  
where
\begin{equation}\label{bbf}
\bar{f}_b(\beta ,\gamma )=\sum_{r=0}^{c'-1}f_b(\beta +r,\gamma ) . 
\end{equation}
On the other hand, using the explicit Eq. (\ref{qesp}) with the $(al')$th iteration of Eq. (\ref{pqep1}), we get
\begin{equation}\label{dcqespen}
\left|\Psi_{b,\beta +c',\gamma}\right\rangle = \exp [ic'\gamma +i\bar{w}_b(\beta ,\gamma )]\left|\Psi_{b,\beta ,\gamma}\right\rangle ,
\end{equation}
where
\begin{equation}\label{bw}
\bar{w}_b(\beta ,\gamma )=\sum_{r=0}^{al'-1}w_b\left(\beta +\frac{rc'}{al'},\gamma\right) 
\end{equation}
and $w_b(\beta ,\gamma )$ are the phases in Eq. (\ref{pqep1}). Then, by comparing Eq. (\ref{dcqespe}) with Eq. (\ref{dcqespen}), it follows that
\begin{equation}\label{comp}
c'\gamma +\bar{w}_b(\beta ,\gamma )=\bar{f}_b(\beta ,\gamma )+2\pi m,
\end{equation}
where $m$ is some integer. Next, let us add $c'$ equations (\ref{ci1}) for $\beta,\beta+1,\dots,\beta+c'-1$. Using Eq. (\ref{bbf}), we find that
\begin{eqnarray}\label{ci1t}
\bar{f}_b\left(\beta, \gamma +\frac{2\pi}{c'}\right) & - & \bar{f}_b(\beta, \gamma ) \\ \notag & + & g_b(\beta, \gamma ) -g_b(\beta +c', \gamma ) =2\pi c'S_b . 
\end{eqnarray}
{\ }\\
Similarly, by adding $al'$ equations (\ref{ci2}) for $\beta,\beta +c'/(al'),\dots,\beta+c'(al'-1)/(al')$ and using Eq. (\ref{bw}), we get 
\begin{eqnarray}\label{ci2t}
\bar{w}_b\left(\beta, \gamma +\frac{2\pi}{c'}\right) & - & \bar{w}_b(\beta, \gamma ) \\ \nonumber & + & g_b(\beta, \gamma )-g_b\left(\beta +c', \gamma \right) =2\pi al'C_b . 
\end{eqnarray}
Because of Eq. (\ref{comp}), Eq. (\ref{ci1t}) can be rewritten as
\begin{eqnarray}\label{ci1tn}
2\pi +\bar{w}_b\left(\beta, \gamma +\frac{2\pi}{c'}\right) & - & \bar{w}_b\left(\beta, \gamma \right) + g_b(\beta, \gamma )\\ \notag & - &  g_b(\beta +c', \gamma ) =2\pi c'S_b . 
\end{eqnarray}  
Finally, by comparing Eq. (\ref{ci2t}) with Eq. (\ref{ci1tn}), we obtain the DE:
\begin{equation}\label{DE}
c'S_b-al'C_b=1.
\end{equation}
Equation (\ref{DE}) determines the allowed values of the integer $C_b$ which characterizes the adiabatic quantum transport in angular momentum of the ODKRs, see previous sections. Clearly, $C_b=0$ in Eq. (\ref{DE}) only in the exceptional case of $c'=1$ (QE spectrum with only one band), so that the DE shows the general topological nontriviality of the ODKR ensemble. We also see, from $l'/c'=l/c$ with coprime integers $(l',c')$, that the integers $a$, $c'$, and $l'$ in Eq. (\ref{DE}) depend quite erratically on $l$ and on the rational value of the parameter $\eta =a/c$, determining both classical and quantum properties of the DKRs \cite{id2}.

\begin{center}
\textbf{VII. EXAMPLES}
\end{center} 

To illustrate some of the concepts above and to verify the validity of Eq. (\ref{DE}), we consider here some examples for which the Chern integers $C_b$ were calculated numerically in work \cite{jg2} and show that these values of $C_b$ satisfy Eq. (\ref{DE}) for integer $S_b$. In the examples, the kicking potentials in Eq. (\ref{HDKR}) are $V(\theta )=\tilde{V}(\theta )=\cos (\theta )$, $l=2$ in Eq. (\ref{orc}), and $\eta\hbar =2\pi /3$. Since $\eta\hbar =2\pi al'/c'$, the latter value means that $al'=1$ and $c'=3$ (thus, $a=l'=1$, so that from $l'/c'=l/c$ and $l=2$ we see that $c=6$). Values of $C_b$ for the $c'=3$ QE bands $b$, calculated in Ref. \cite{jg2} for different values of $K/\hbar$ and $\tilde{K}/\hbar$ (see Figs. 5 and 7 in Ref. \cite{jg2}), are listed in Table I with the corresponding values of $S_b$ from Eq. (\ref{DE}). As all the values of $S_b$ are integers, this shows the validity of Eq. (\ref{DE}).
\begin{table}[ht!]
\caption{See text for details. }
\begin{tabular}{||c||c|c|c|c||}
\hline
$K/\hbar$ & 2 & 6 & 8 & 1 \\
\hline
$\tilde{K}/\hbar$ & 2 & 6 & 8 & 5 \\
\hline
$C_1$,\ $S_1$ & -1,\ 0 & -4,\ -1 & -16,\ -5 & 2,\ 1 \\
\hline
$C_2$,\ $S_2$ & 2,\ 1 & 8,\ 3 & 32,\ 11 & -4,\ -1 \\
\hline
$C_3$,\ $S_3$ & -1,\ 0 & -4,\ -1 & -16,\ -5 & 2,\ 1 \\
\hline
\end{tabular}
\label{tab:}
\end{table}

From Eq. (\ref{QEBZ1}), one expects the $c'=3$ QE bands $\omega_b^{(\alpha )}(\gamma )$ to be periodic in $\alpha$ with period $2\pi /c'=2\pi /3$, This is clearly shown in Fig. 6 of Ref. \cite{jg2}.

\begin{center}
\textbf{VIII. SUMMARY AND CONCLUSIONS}
\end{center}

The ODKR ensemble \cite{jg1,jg2}, studied in this paper using the generalized version (\ref{UODKR}) of its evolution operators, seems to be the first topologically nontrivial Floquet system for which its Chern integers $C_b$ can be given an \emph{exact} quantum-transport interpretation: As shown in Ref. \cite{jg1}, $C_b$ exactly determines, on a first-principles basis, the change in the mean angular momentum of a state in band $b$ when the parameter $\alpha$ in Eq. (\ref{UODKR}) is adiabatically varied from $0$ to $2\pi$. Such an exact interpretation should be contrasted with the \emph{approximate} one, based on linear-response theory, given to a TKNN Chern integer, as determining the quantum Hall conductance carried by a magnetic band for the system of Bloch electrons in a uniform magnetic field \cite{tknn}. On the other hand, the phase-plane translational invariance of the latter system allows one to derive a well-known and useful Diophantine equation (DE) for the TKNN integers \cite{tknn,daz,dz} and one may ask whether a similar equation exists also for the ODKR-ensemble integer $C_b$. 

In this paper, we have answered this question in the affirmative under most general conditions. This was done in several stages. First, we have introduced a DKP system described by the evolution operator (\ref{UP}) or (\ref{UPn}), which is translationally invariant in the position-momentum phase plane. We have then established a correspondence between this system and the ODKR ensemble, see Eqs. (\ref{eqr}), (\ref{eqrqe}), and (\ref{eqrc}). Next, this correspondence was used to derive several results concerning the topological and related properties of the QE band spectrum and eigenstates of both the DKP system and the ODKR ensemble. 

Among these results, the main one is the DE (\ref{DE}) determining the allowed values of two Chern integers, $C_b$ and $S_b$, where $C_b$ characterizes the adiabatic quantum transport of ODKRs \cite{jg1}, see above. The second integer, $S_b$, is associated with the DKP system and appears in this paper apparently for the first time. The quantities $al'$ and $c'$ featured by the DE (\ref{DE}) have a well-defined meaning: They give the degeneracy of the QE band eigenstates of the DKP system and of the ODKR ensemble, respectively (see end of Sec. V).

While both Chern integers in the DE for the TKNN system have a quantum-transport interpretation \cite{tknn,daz,kunz}, the physical meaning of the second Chern integer $S_b$, associated with our DKP system, is not yet known. We plan to investigate this problem in future works. 

\begin{center}
\textbf{APPENDIX A}
\end{center}

Since the eigenvalues of $\hat{L}$ in Eq. (\ref{UDKR}) are $n\hbar$ for all integers $n$, one has $\hat{L}^{2}/(2\hbar )=\pi l n^2$ identically under the condition (\ref{orc}). This implies, for even $l$, that $\exp [-i\hat{L}^{2}/(2\hbar )]=1$. The evolution operator (\ref{UDKR}) reduces then to the ODKR operator (\ref{UODKR}).  

Let us require the operator (\ref{UODKR}) to be invariant under translations in angular momentum $\hat{L}$ by $N\hbar$, for some integer $N$. After replacing $\hat{L}$ by $\hat{L}+N\hbar$ in Eq. (\ref{UODKR}), there will appear factor terms $\exp (\pm iN\eta \hat{L})$. Since $\hat{L}=-i\hbar d/d\theta$, these terms are equal to $\exp (\pm N\hbar\eta d/d\theta )$, i.e., translations of $\theta$ by $\pm N\hbar\eta$. One then finds that $\hat{U}_{\rm ODKR}^{(\alpha )}(\theta,\hat{L}+N\hbar)=\hat{U}_{\rm ODKR}^{(\alpha +N\hbar\eta)}(\theta )$. Because of the $2\pi$-periodicity of the potentials, it follows that $\hat{U}_{\rm ODKR}^{(\alpha )}(\theta,\hat{L}+N\hbar)$ will be equal to $\hat{U}_{\rm ODKR}^{(\alpha )}(\theta,\hat{L})$ only if $N\hbar\eta$ is a multiple of $2\pi$. This and Eq. (\ref{orc}) imply that $\eta$ is rational, i.e., Eq. (\ref{eta}). Also, writing $l/c=l'/c'$, where $l'$ and $c'$ are coprime integers, we see that the minimal value of $N$ for which $N\hbar\eta$ is a multiple of $2\pi$ is $N=c'$, i.e., Eq. (\ref{N}).


\begin{thebibliography}{99}
\bibitem{note} For the ordinary KHM \cite{qc1,qc2,qc3,qc4}, $W_p$ and $W_x$ are proportional to cosine functions.
\bibitem{note1} One can easily show that the $c'$ degenerate states $\Phi_{b,\gamma }^{(\alpha +2\pi s/c')}(n)$, $s=0,1,...,c'-1$, are essentially the shifted states $\Phi_{b,\gamma }^{(\alpha)}(|n+r|_{c'})E_{n+r}(\gamma )$, $r=0,1,...,c'-1$, where $|n+r|_{c'}$ means $n+r$ mod($c'$) and $E_{n+r}(\gamma )=\exp[ic'\gamma (|n+r|_{c'}-n-r)]$ is a Bloch factor emerging from a translation of $n$ by $c'$, see Eq. (\ref{qes}). Due to the invariance under the latter translation, the $c'$ states above may be viewed as representations of the same state.
\end{thebibliography}
\end{document}